\documentclass{aa}
\usepackage{graphicx}
\usepackage{times,float}

\begin{document}

\title{Detection of an optical transient following the 13 March 2000 
short/hard gamma-ray burst
      \thanks{Based in part on observations made with the BOOTES instruments in
 South Spain.}
      }

   \author{A.J. Castro-Tirado
          \inst{1,2}
   \and J.M. Castro Cer\'on
          \inst{3}
   \and J. Gorosabel
          \inst{1,2,4}
   \and P. P\'ata
          \inst{5}
   \and J. Sold\'an
          \inst{6}
   \and R. Hudec
          \inst{6}
   \and M. Jelinek
          \inst{6}
   \and M. Topinka
          \inst{6}
   \and M. Bernas
          \inst{5} 
   \and T.J. Mateo Sanguino
          \inst{7}
   \and A. de Ugarte Postigo
          \inst{8}
   \and J.\'A. Bern\'a
          \inst{9}
   \and A. Henden
          \inst{10,11}
   \and F. Vrba
          \inst{11}
   \and B. Canzian
          \inst{11}
   \and H. Harris
          \inst{11}
   \and X. Delfosse
          \inst{12}
   \and B. de Pontieu
          \inst{13}
   \and J. Polcar
          \inst{14} 
   \and C. S\'anchez-Fern\'andez
          \inst{2}
   \and B.A. de la Morena
          \inst{7}
   \and J.M. M\'as-Hesse
          \inst{2}
   \and J. Torres Riera
          \inst{15}
   \and S. Barthelmy
          \inst{16}
          }

\offprints{A. J. Castro-Tirado, \\
    \email{ajct@iaa.es}
    }

   \institute{Instituto de Astrof\'{\i}sica de Andaluc\'{\i}a (IAA-CSIC), 
              P.O. Box 03004, 18080 Granada, Spain 
   \and Laboratorio de Astrof\'{\i}sica Espacial y F\'{\i}sica 
              Fundamental (LAEFF-INTA), 28080 Madrid, Spain
   \and Real Instituto y Observatorio de la Armada, Secci\'on de 
              Astronom\'{\i}a, 11.110 San Fernando--Naval (C\'adiz) Spain
   \and Danish Space Research Institute, Juliane Maries Vej 30, 2100 Copenhagen \O\ Denmark.
   \and Czech Technical University, Faculty of Electronic Engineering, 
        Department of Radioelectronics, 166 27 Prague, Czech Republic 
   \and Astronomical Institute of the Czech Academy of Sciences,
              251 65 Ond\v{r}ejov, Czech Republic
   \and Centro de Experimentaci\'on del Arenosillo (CEDEA-INTA),
              E-21130 Mazag\'on, Huelva, Spain 
   \and Departamento de Astrof\'{\i}sica, Universidad Complutense, 
              Madrid, Spain
   \and Departamento de F\'{\i}sica, Ingenier\'{\i}a de Sistemas y 
        Teor\'{\i}a de la Se\~nal, Universidad de Alicante, Alicante, Spain
   \and Universities Space Research Association, Flagstaff station, AZ, U.S.A.
   \and U. S. Naval Observatory, Flagstaff station, AZ, U.S.A.
   \and Laboratoire d\' \rm Astrophysique de l\' \rm Observatoire 
        de Grenoble, Grenoble, France
   \and Lockheed Martin Solar \& Astrophysics Lab, Palo Alto, 3251 Hanover 
        St., Bldg. 252, CA 94304, U.S.A.
   \and Dept. of  Theoretical Physics and Astrophysics, Masaryk University, 
        Brno, Czech Republic
   \and Divisi\'on de Ciencias del Espacio (DCE-INTA), 
        Torrej\'on de Ardoz, E-28850 Madrid, Spain 
   \and NASA Goodard Space Flight Center, Greeebelt, MA, U.S.A.
             }

\date{Received / Accepted }

\abstract{ We imaged the error box of a gamma-ray burst of the short (0.5 s), 
    hard type (GRB 000313), with the BOOTES-1 experiment in southern Spain, 
    starting 4 min after the $\gamma$--ray event, in the $I$-band. 
    A bright optical transient (OT 000313) with $I$ = 9.4 $\pm$ 0.1 was found 
    in the BOOTES-1 image, close to the error box (3$\sigma$) provided 
    by BATSE. Late time $VRIK^\prime$-band deep observations failed to reveal 
    an underlying host galaxy. If the OT 000313 is related to the short, 
    hard GRB 000313, this would be the first optical 
    counterpart ever found for this kind of events (all counterparts to date 
    have been found for bursts of the long, soft type). The fact that only 
    prompt optical emission has been detected (but no afterglow emission at 
    all, as supported by theoretical models) might explain why no optical 
    counterparts have ever been found for short, hard GRBs.
    This fact suggests that most short bursts might occur in a low-density 
    medium and favours the models that relate them to binary mergers in very 
    low-density environments.

\keywords{gamma rays: bursts -- optical transients -- techniques: photometric -- cosmology: observations}
         }

\maketitle

\section{Introduction}

Gamma Ray Bursts (GRBs hereafter) are flashes of cosmic high energy photons, 
and they remained for 25 years one of the most elusive mysteries for high 
energy astrophysicists, the main problem being the lack of knowledge about 
the distance scale. The detection of counterparts at other wavelengths for 
the long duration, soft GRBs, revealing their cosmological origin (see van 
Paradijs et al. 2000 for a recent review). Thus, counterparts 
to about 30 bursts have been discovered so far with about 25 redshifts 
measured, but all of them belong to the so called long duration 
($\sim$ 20 s), soft bursts class that comprises about 75\% of all GRBs 
(Mazets et al. 1981). 
There are evidences that the two classes of bursts are different: whereas 
long bursts have softer spectra, short bursts have harder spectra 
(Dezalay et al. 1996). The latter ones comprise about 25\% of all GRBs 
(Kouvelioutou et al. 1993) and their origin still remain a puzzle. No 
counterparts at longer wavelengths have been found yet in spite of intense 
efforts in order to detect the optical, infrared and radio counterparts to 
several short, hard bursts (Kehoe et al. 2001, Gorosabel et al. 2002, 
Hurley et al. 2002, Williams et al. 2002). 
Therefore, one of the remaining GRB mysteries is whether the origin of 
the two populations are substantially different from one another.

\begin{figure}[t]
\begin{center}
  \resizebox{8cm}{4.1cm}{\includegraphics{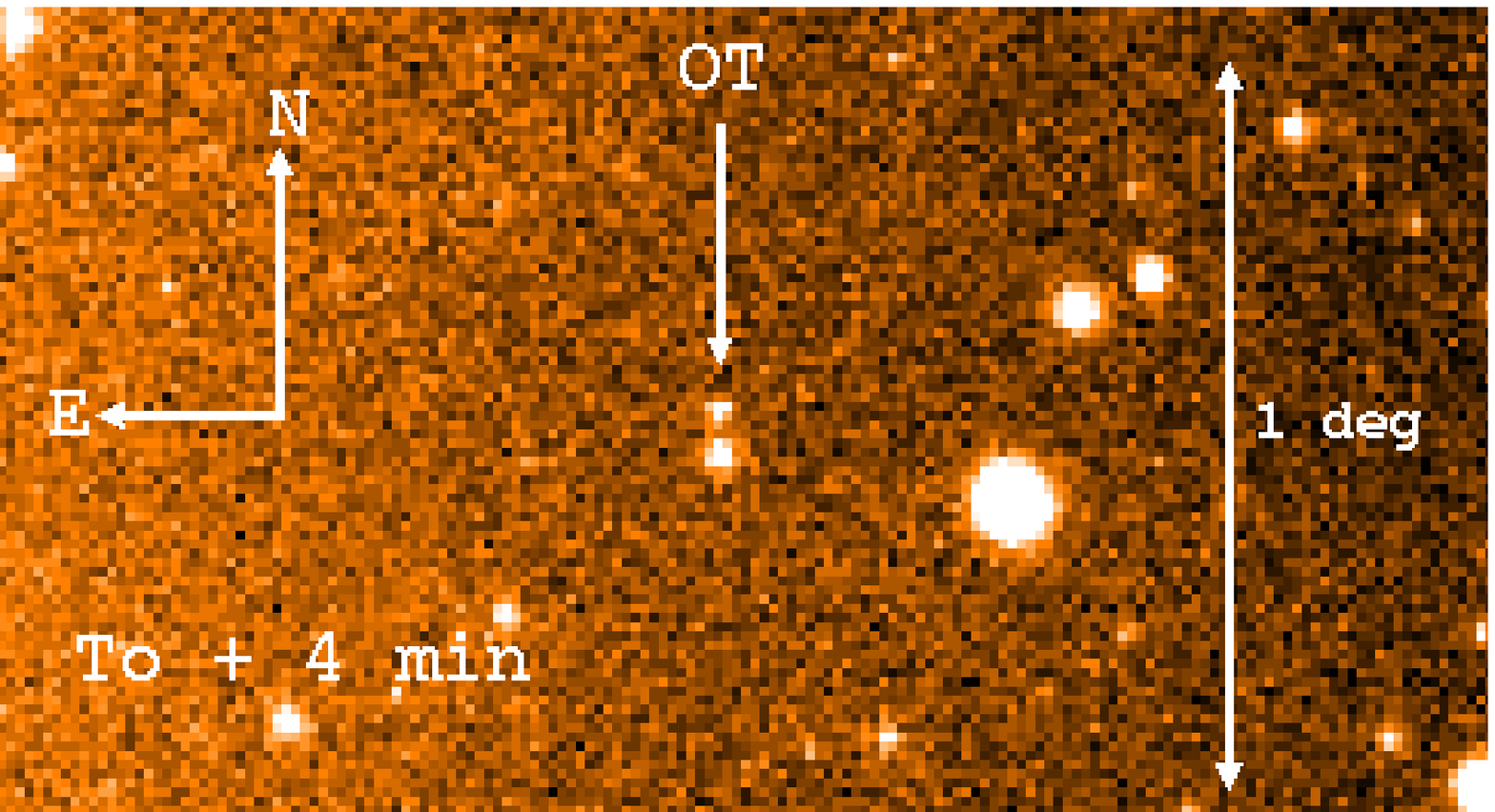}}
  \resizebox{8cm}{4.1cm}{\includegraphics{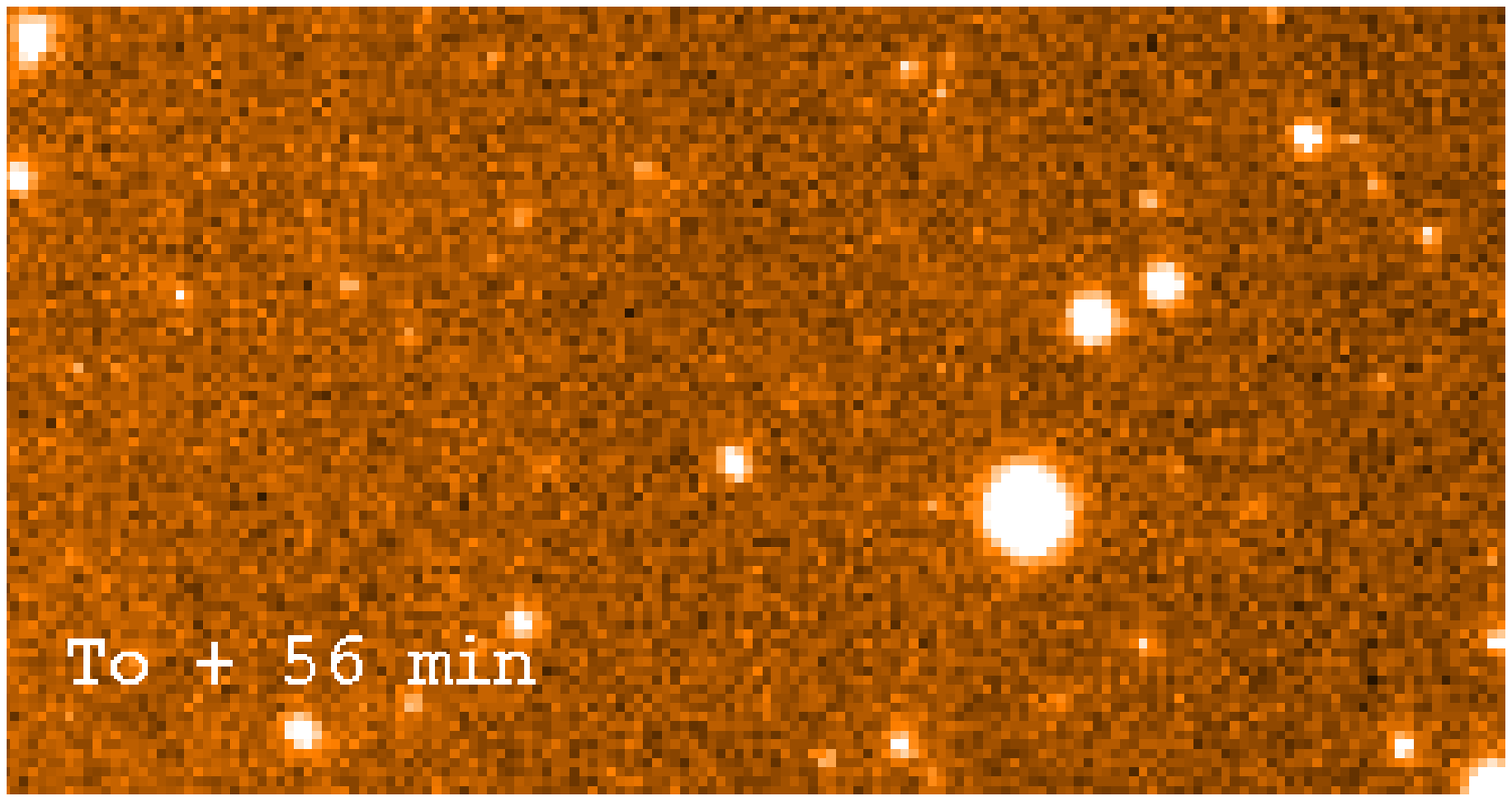}}
    \caption{$2.1^{\circ} \times 1.1^{\circ}$ fields in the $I$-band 
     containing a 
     fraction of the BATSE GRB error box. 13 Mar, UT 21 h 17 min 
     ({\it upper panel}) and 13 Mar, UT 22 h 13 min
     ({\it left panel}). The position of the OT 000313 is marked 
     with an arrow at the center of the image.
     North is upward and East is to the left. The limiting magnitude is 
     $I$ = 12.0 for the second image.}
\end{center}
\end{figure}

Here we present the results of a follow-up observation 
for one of these short/hard events.
GRB 000313 was detected on 13 March 2000, UT 21 h 13 min 04 s by the
Burst and Transient Source Experiment (BATSE) instrument aboard the 
Compton Gamma-ray Observatory (trigger number 8035). 
The  single--peaked gamma--ray burst lasted 0.768 $\pm$ 0.458 s, 
showed substructure and was possibly 
detected below 50 keV. It reached a peak flux (50-300 keV)
of $\sim$ 1.2 ($\pm$ 0.1) $\times$ 10$^{-7}$ erg cm$^{-2}$ s$^{-1}$
and a fluence ($\geq$ 25 keV) of $\sim$ 2.6 ($\pm$ 1.7) $\times$ 10$^{-7}$ 
erg cm$^{-2}$. The gamma-rays properties as well as the duration of the burst 
make from GRB~000313 a clear short/hard gamma-ray burst.
The original BATSE position reported through the GCN/BACODINE Network 
(Barthelmy et al. 1998) was
R.A.(2000) = 13h 31m, Dec(2000) = +27$^{\circ}$ 10$^{\prime}$
which was later on refined to 
R.A.(2000) = 13h 11m, Dec(2000) = +10$^{\circ}$ 14$^{\prime}$.

\section{Observations}
 \label{observaciones}

We obtained $I$-band and unfiltered images with the wide field CCD cameras of
the Burst Observer and Optical Transient Exploring System (BOOTES-1) 
(Castro-Tirado et al. 1999a) beginning 4 min after the event 
(13 Mar, UT 21 h 17 min).
The image taken with the ultrawide field CCD camera (a 1,524 x 1,024 pixel CCD 
attached to a 18-mm f/2.8 lens yielding 1$^{\prime}$.64/pixel) covered the 
original BATSE error box whereas a mosaic of 9 images was performed with the
wide field CCD camera (a 1,524 x 1,024 pixel CCD attached to a 50-mm f/2.0 
lens yielding 0$^{\prime}$.64/pixel) in order to cover the full error box. 
The narrow field
CCD camera at the Cassegrain focus of the 0.3-m BOOTES-1 telescope (yielding 
2$^{\prime\prime}$/pixel) imaged
part of the field starting on 14 March, 00 h 05 min (4.8 hr later),
once the OA was confirmed by the BOOTES team.
Further observations were made the same night with the CCD camera at the 
1.0-m Jacobus Kapteyn Telescope (JKT) at La Palma under very poor seeing 
conditions and with the 1.55-m 
Telescope at the U.S. Naval Observatory (USNO) in Flagstaff. Late time 
observations were obtained with the MAGIC NIR camera on the 1.23-m telescope
at Calar Alto (CAHA) and with the ALFOSC instrument on the Nordic Optical 
Telescope (NOT) at La Palma. Table 1 report the list of observations.
All frames were de-biased and flat-fielded using standard procedures.

\begin{figure}
\resizebox{\hsize}{!}{\includegraphics[angle=0]{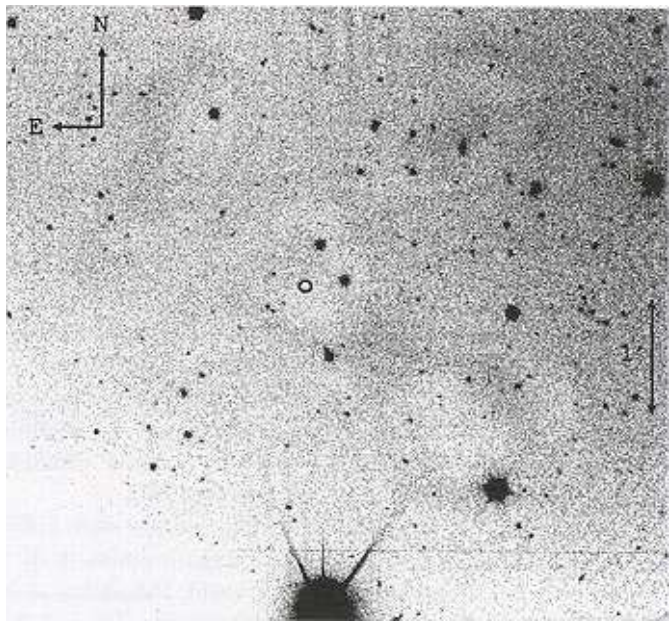}}
    \caption{The deep $I$-band image obtained at the position of the OT
000313 with the NOT on 31 Mar 2001. The total integration time was
1,800--s. The circle shows the 3$^{\prime\prime}$ radius error box for 
the OT derived from the BOOTES-1 image. The bright star at the bottom is
the star 4' south of the OT in Figure 1. The field is 2.9' $\times$ 2.1'  
with North upward and East to the left.}
\label{figure2}
\end{figure}

\section{Results}

A comparison of the frames acquired by the BOOTES-1 wide field CCD 
on Mar 13, 21 h 17 min UT and Mar 13, 22 h 09 min UT 
revealed an optical transient (OT) (see Figure 1) at R.A.(2000) = 
13h 50m 07.9s, Dec(2000) = +31$^{\circ}$ 16$^{\prime}$ 49$^{\prime\prime}$ 
($\pm$ 3$^{\prime\prime}$) (Castro-Tirado et al. 2000).
The object is not detected in the rest of BOOTES-1 images
covering the OT position that were taken during the night, starting at 
22 h 09 min UT.
Using aperture photometry software, we could determine the magnitude of the
optical transient in the image as $I$ = 9.4 $\pm$ 0.1, by comparison with 
secondary standards in the field (Henden 2000). The object is not
present in the simultaneous ultrawide field CCD frame taken with the
18mm lens at 21 h 17 min UT, but only an upper limit of $R$ $>$ 9.1 can be 
derived, which explains the non-detection.
Late-time observations, carried out between 40 days and $\sim$ 2 year later, 
have failed to reveal any quiescent source within the OT error circle down 
to $V$ $\sim$ 23.5, $R$ $\sim$ 24.5, $I$ $\sim$ 23.5 and $K$' $\sim$ 18.0 
(Figure 2).

\begin{table}[bt]
\begin{center}
\caption{Log of the OT 000313 optical and near-infrared follow-up observations}
  \begin{tabular}{cccccc}
\hline
Date    & Time      & Telescope & Filter & Exposure & Magnitude  \\
        &  (UT)     &           &        & time (s) &            \\
\noalign{\smallskip}
\hline
\noalign{\smallskip}
13 Mar 00 & 21:17  & 0.05 BOO     &  $I$  &   300   & 9.4$\pm$0.1 \\
13 Mar 00 & 22:13  & 0.05 BOO     &  $I$  &   300   &  $>$ 12.0 \\
14 Mar 00 & 02:02  & 0.3 BOO      &  $-$  & 9 x 120 &  $>$ 19.0 \\
14 Mar 00 & 04:00  & 1.0 JKT      &  $B$  &   600   &  $>$ 20.1 \\
14 Mar 00 & 04:12  & 1.0 JKT      &  $V$  &   600   &  $>$ 19.3 \\
14 Mar 00 & 04:25  & 1.0 JKT      &  $R$  & 2 x 300 &  $>$ 19.0 \\
14 Mar 00 & 04:35  & 1.0 JKT      &  $I$  & 2 x 300 &  $>$ 18.0 \\
14 Mar 00 & 05:11  & 1.5 USNO     &  $I$  &    60   &  $>$ 20.2 \\
14 Mar 00 & 06:16  & 1.5 USNO     &  $R$  &   300   &  $>$ 21.4 \\
14 Mar 00 & 07:19  & 1.5 USNO     &  $I$  &   900   &  $>$ 21.6 \\
23 Apr 00 & 22:00  & 1.2 CAHA     &  $K'$ &  2,400  &  $>$ 18.0 \\
28 Apr 00 & 00:00  & 2.5 NOT      &  $I$  & 3 x 600 &  $>$ 22.4 \\
22 Mar 01 & 02:20  & 2.5 NOT      &  $V$  & 3 x 750 &  $>$ 23.4 \\
30 Mar 01 & 00:00  & 2.5 NOT      &  $R$  & 4 x 900 &  $>$ 24.0 \\
31 Mar 01 & 22:00  & 2.5 NOT      &  $I$  & 6 x 300 &  $>$ 23.5 \\
20 May 02 & 02:00  & 2.5 NOT      &  $R$  & 15 x600 &  $>$ 24.5 \\
\hline
\end{tabular} 
\end{center}
\end{table}

\section{Discussion}

\subsection{The reality of the object}

The  OT is point-like,  with the  same point-spread-function  (PSF) as
other field  stars.  The PSF of  the OT image (and  also of comparison
stars  in  the field)  was  fitted  with  a two  dimensional  Gaussian
function,    making    use     of    the    nonlinear    least-squares
Marquardt-Levenberg  algorithm. From  the fitted  profiles  it follows
that the OT  it a real star-like object.  We can  also exclude a glint
of  a  satellite (for  example  a  Molniya  satellite, i.e.   with  an
inclination larger than 60$^{\circ}$) because the image is not trailed
in spite  of being exposed for  300 s, as  seen is many of  the BOOTES
database images.  Moreover, a search on satellite  databases has given
negative results.  During this time  frame, within 0.5 degrees of this
position, there was only one  satellite, namely NORAD no. 14,473, this
is  a small piece  of debris  of a  rocket launch  with a  radar cross
section  of 0.05  m$^{2}$.   It appears  on  a track  that remains  at
0.5$^{\circ}$ away from the measured  position of the OT and the track
it  follows in  the  sky never  gets  closer than  0.5 degrees.   More
importantly, this  satellite is tiny, and  is at a range  of $>$ 3,000
km.   Even under  full sun  conditions, magnitude  10 is  difficult to
believe for  this object.   In reality, the  orbit is deeply  into the
Earth's  shadow,  almost  1,000   km.   Under  such  conditions,  only
moon-light  could reflect  off of  it, which  renders it  at  least 13
magnitudes fainter, and definitely fainter  than mag 18.  So, under no
conditions can the optical transient  be this satellite.  There are no
other known candidate satellites to explain it.  In addition, the tiny
angular extent of the  observed optical signal (0.03$^{\circ}$), tells
that for this to be a satellite, it would have to have been an optical
glint or flash of {\it very short} duration ($<$ 0.25 s).  An airplane
flash is  ruled out  as no other  such event  is detected in  the full
16$^{\circ}$  x 11$^{\circ}$  field of  view.  We  can also  exclude a
cosmic-ray (CR) mimicking  an OT as the mean rate of  CR in the BOOTES
ST-8 CCD  cameras is  0.1 min$^{-1}$ cm$^{-2}$,  i.e. 0.06 in  a 300-s
typical exposure.  The  probability of having a CR  with a PSF similar
to  that  of a  star  in  the field  is  $\leq$  10$^{-2}$, thus  this
possibility  can  be  significantly  reduced ($P$  $\leq$  6  $\times$
10$^{-4}$).   A  head-on  meteor can  be  also  ruled  out, as  it  is
extremely improbable,  of order of  $<$ 10$^{-6}$ (Hudec  1993, Varady
and Hudec  1992).  Moreover, the PSF analysis  indicates no detectable
trailing for the image.

\subsection{A relationship to GRB 000313?}

The OT 000313 is 23$^{\circ}$ from the center of the refined BATSE error box
(the so-called Hunstville position). 
The statistical-only error radius is 7.6$^{\circ}$, and  
the total error is best described by a two-component model, sum of two 
Gaussians, Briggs et al. 1999).
Thus, the OT lies at 3.0$\sigma$ from the center of the refined GRB 000313.
The fact that this event was not detected by any other satellite resulted
in the lack of a more accurate position from the high-energy data itself.

In the  GRB fireball model  the prompt optical  flash seen in  GRBs is
thought  to arise  when a  reverse shock  propagates into  the ejected
shell  (Sari and  Piran  1999), whereas  the  afterglow emission  that
starts few minutes after the event is thought to be due to the forward
shock propagating  into the interstellar medium (ISM)  after the shell
has  swept up  a considerable  volume of  the ISM  that  surrounds the
central  engine (M\'esz\'aros  and Rees  1997).  If  the OT  000313 is
related to  the GRB 000313  and we assume  a power-law decay  with the
flux  $F$ $\propto$  t$^{-\alpha}$ then,  the derived  power-law decay
index  is $\alpha$  $\geq$ 2.2,  only comparable  to the  steepest GRB
optical  afterglows (Castro-Tirado et  al. 2001).  Figure 3  shows the
light curve  of the OT 000313  superimposed to the light  curve of the
GRB  990123  event for  which  a  prompt  optical flash  was  detected
(Akerlof et al.  1999). The OT 000313 data are  consistent with a fast
decay  similar to  that  of GRB  990123  ($\alpha$ =  2.1, Akerlof  et
al. 1999) but with the  absence of the optical afterglow (that started
at  about  0.01 day  after  the occurrence  of  GRB  990123).  We  can
interpret this observational fact considering that only prompt optical
emission has been  detected in the OT 000313  but no optical afterglow
emission  at all.   No radio  afterglow emission  was  detected either
(Berger et al.  2000, Frail 2000), just like none  has ever been found
in other short, hard GRBs (Gorosabel et al. 2002, Hurley et al. 2002).

\begin{figure}
\resizebox{\hsize}{!}{\includegraphics[angle=0]{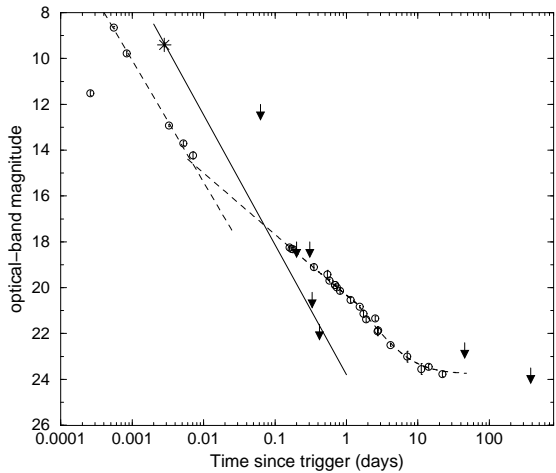}}
    \caption{The light curve of the OT 000313 ($I$-band, star, upper 
     limits and solid line) superimposed to the light curve of the long-
     duration GRB 990123 ($R$-band, empty circles and dashed line), the 
     only burst for which a prompt optical flash has been detected. The 
     OT 000313 data are consistent with a prompt optical flash fast decay 
     ($\alpha$ $\geq$ 2.2) comparable to that of the GRB 990123 
     ($\alpha$ = 2.1) but with the absence of an optical afterglow 
     (that started at about T$_0$ + 0.01 day in the GRB 990123). The GRB 
     990123 data are taken from Akerlof et al. (1999), Castro-Tirado 
     et al. (1999b) and references therein.}
\label{figure2}
\end{figure}

Although the origin of the long duration, 
soft GRBs seems to be widely accepted as the collapse of massive stars 
(Woosley 1993), the origin of the short 
duration, hard GRBs is still an open question. It has been proposed that the 
extremely brief bursts ($<$ 100 ms) might be due to primordial black hole 
(BH) evaporations (Hawking 1974, Cline et al. 1999) although most of the 
short, hard burst 
population is thought to be due to binary mergers (Narayan et al. 1992). 
Lifetimes of neutron star--neutron star (NS) systems are of the 
order of $\sim$ 10$^9$ yr, and large escape velocities are usual, putting 
them far away from the regions where their progenitors were born. Therefore, 
the GRB progenitors in the binary merger model context would be located in 
very low density regions, where no afterglow emission is expected. The 
likely result is a Kerr black hole and the energy released during the merging 
process is $\sim$ 10$^{54}$ erg, provided that the disk is sufficiently small 
and the accretion is driven by neutrino cooling (Narayan et al. 2001). In 
that case, the expected duration of the relativistic wind (i.e. the GRB) is 
$\sim$ 1 s. A similar ``output'' would be expected from a NS-BH merger 
(Paczy\'nski 1991).

Theoretical models have recently claimed that short GRBs only could
exhibit very faint optical afterglow  emission (R $>$ 23, a few hours after
the gamma-ray event, Panaitescu et al. 2001), therefore consistent with 
our upper limits.
Although most NS--NS  mergers should take place within a few tens of
kpc from their host galaxies (see Fig. 21 of Fryer, Woosley and Hartmann 1999
for a range of masses of galaxies), the fact that no host galaxy is found 
within the error circle down
to the above mentioned upper limits is not unusual, due to the fact that
most host galaxies are fainter than $R$ = 24 (R = 24.8 is the median 
apparent magnitude, Djorgovski et al. 2001).

If the OT 000313 is related to the short GRB 000313 then, the
fact that only prompt optical emission has been observed (and no
optical afterglow emission) might explain the fact that no other 
optical counterpart for the short GRB class has been detected.
These observational facts might indicate that 
short GRBs occur in a low-density medium, favouring the models that 
relate them to binary mergers in galactic haloes
or in the intragalactic medium.

We however note that a low density medium also makes it difficult to produce 
a 9 mag optical flash by the reverse shock, unless the bulk Lorentz factor 
$\Gamma$ of the shell would be very large due to one of the following reasons: 
i) the typical reverse shock frequency $\nu_m$ $\propto$ $\Gamma$ $^{2}$ 
n$^{1/2}$ with $n$ the density of the ISM in the surroundings of the GRB. 
A small $n$ requires a 
large $\Gamma$ ($>$ 10$^{2}$) to make the spectrum peaking at optical 
(Kobayashi 2000); 
ii) F$_{\nu,max}$ also positively related with $n$, $\Gamma$ and the 
isotropic energy release E$_{\rm iso}$. 
For short bursts, if both $n$ and E$_{\rm iso}$ are small compared 
with the long bursts, $\Gamma$ should be very large to compensate the deficit; 
iii) to achieve a bright optical flash, the shell should be either thin 
($\Delta_{0}$ $<$ $l$/$\Gamma^{8/3}$ with $\Delta_{0}$ the width
of the shell and $l$ the Sedov length i.e., the reverse shock would be 
not relativistic) or rather marginal (even better, i.e. the 
ratio $l$/2$\Delta_{0}$$\Gamma^{8/3}$ $\sim$ 1).

\section{Conclusions}
  \label{conclusiones}
  
If the OT 000313 is indeed related to the short GRB 000313 then, the
fact that only prompt optical emission has been observed (and no
optical afterglow emission) might explain the fact that no other 
optical counterparts for the short GRB class has been detected, 
favouring the models that relate them to binary mergers in galactic 
haloes or in the intragalactic medium. Given the fact that the distance 
distribution of NS-NS mergers depends on the mass of the host galaxy 
(Fryer, Woosley and Hartmann 1999), deeper observations of the OT 000313 
error box might help to better constraint the distance and/or the mass
of the host galaxy.

\begin{acknowledgements}

We  thank the  anonymous referee  and  D.  Hartmann  for the  valuable
suggestions in order to improve  the manuscript.  We are very grateful
to R. Vanderspek  for his independent studying of  the BOOTES-1 image,
to R. M. Kippen for fruitful  conversations on the errors in the BATSE
position and  to G.  J. Fishman  and V.  Connaughton  for checking the
BATSE data on this particular burst.  We also thank M. Ir\'{\i}bar, I.
D\'{\i}az Pe\~na, J. J.  Mart\'{\i}n Franc\'{\i}a, F. Soubrier, J.  M.
Vilaplana, J. A.  Ad\'amez and  F.  Ramos Mantis for their hospitality
and help at the BOOTES-1 station and to J.  Masegosa, A.  del Olmo, L.
Verdes-Montenegro,  L.  Christensen,  J.   Hjorth, B.   Jensen and  H.
Pedersen for their help regarding the NOT observations.  This work has
been  (partially) supported by  the Space  Sciences Division  at INTA,
through  the   project  IGE  4900506,  by  the   Spanish  CICYT  grant
ESP95-0389-C02-02  and  through the  project  GV01-361  of Oficina  de
Ciencia  i  Tecnologia  de   la  Generalitat  Valenciana.   The  Czech
Participation  is supported  by the  Grant Agency  of the  Czech Rep.,
grant  205/99/0145,  and  by   the  ESA  Prodex  Contract  14527.   JG
acknowledges  the receipt  of a  Marie Curie  Research Grant  from the
European Commission.  The Jacobus  Kapteyn Telescope is operated by the
Royal  Greenwich  Observatory  and  the Nordic  Optical  Telescope  is
operated  on the  island  of  La Palma  jointly  by Denmark,  Finland,
Iceland, Norway and  Sweden, in the Spanish Observatorio  del Roque de
los Muchachos  of the Instituto  de Astrof\'{\i}sica de  Canarias. The
NOT  data  have  been taking  using  ALFOSC,  which  is owned  by  the
Instituto de Astrof\'{\i}sica de  Andaluc\'{\i}a (IAA) and operated at
the Nordic  Optical Telescope under agreement between  IAA and NBIfAFG
of the Astronomical Observatory of Copenhagen.  This research made use
of J-track, a software  developed by the Mission Operations Laboratory
at NASA's Marshall Space  Flight Center, to track satellites positions
in the sky.

\end{acknowledgements}

\end{document}